\documentclass[aps,twocolumn,superscriptaddress]{revtex4}
\usepackage{graphicx}
\usepackage{bm}
\usepackage{amsmath}
\usepackage{amsfonts}
\usepackage{amssymb}
\usepackage{psfig}

\begin{document} \title{Polariton quantum blockade in a photonic dot}
\author{A. Verger}
 \affiliation{Laboratoire Pierre Aigrain,
Ecole Normale  Sup\'erieure, 24, rue Lhomond, 75005 Paris, France}
\author{C. Ciuti}
\email{ciuti@lpa.ens.fr} \affiliation{Laboratoire Pierre Aigrain,
Ecole Normale  Sup\'erieure, 24, rue Lhomond, 75005 Paris, France}
\author{I. Carusotto}
\affiliation{BEC-CNR-INFM and Dipartimento di Fisica, Universit\`a
di Trento, I-38050 Povo, Italy}

\begin{abstract}
We investigate the quantum nonlinear dynamics of a resonantly
excited photonic quantum dot embedding a quantum well in the
strong exciton-photon coupling regime. Within a master equation
approach, we study the polariton quantum blockade and the
generation of single photon states due to polariton-polariton
interactions as a function of the photonic dot geometry, spectral
linewidths and energy detuning between quantum well exciton and
confined photon mode. The second order coherence function
$g^{(2)}(t,t')$ is calculated for both continuous wave and pulsed
excitations.
\end{abstract}
\date{\today} \maketitle
Several recent developments in the field of quantum
information~\cite{Zeilinger} and quantum
communication~\cite{Gisin} are based on light beams with strongly
non-classical properties. Many techniques have been developed to
obtain such beams, using, e.g.,  parametric down-conversion
processes in bulk nonlinear crystals~\cite{Zeilinger}, colored
centers in diamond ~\cite{Grangier_1} or by taking advantage of
semiconductor electronic quantum
nanodots~\cite{JM_Gerard,Imamoglu_3,Yamamoto}. These are a sort of
artificial two-level atoms, which are able to absorb and diffuse
one quantum of radiation at a time, so that the emitted light
shows strong antibunching properties and a train of single-photon
pulses can be obtained under pulsed excitation. However, the use
of self-organized quantum nanodots requires non-trivial
nanotechnology to control the emission frequency and spatial
position\cite{Imamoglu_2}. In addition to this, the coupling to
the photon mode is far from optimal, due to the large mismatch
between the spatial size of the electron nanodot and of the
photonic mode, as attested by the intrinsically small vacuum Rabi
energy\cite{QD_strong,Tejedor} (typically a fraction of a meV).

Quantum wells strongly coupled to planar microcavities combine
strong nonlinearities due to exciton-exciton interactions with an
efficient coupling to radiation. In particular, their use as
parametric amplifiers and oscillators working at low pump
intensities appears promising
~\cite{Savvidis_00,Cristiano_R_00,Saba_01,review_Cristiano,review_Deveaud,Iac_Crist_1}.
Very recently, lithographic techniques have been developed to
create high quality photonic dots able to confine the photon
without spoiling the strong-coupling with the quantum well
exciton. In this way, polaritons result confined in all three
dimensions with large vacuum Rabi splittings (several
meVs)~\cite{Gregor_02,EPFL}.

In this Letter, we discuss a proposal of a single photon source
based on such a kind of polariton quantum dot as the active
medium. If the photonic confinement volume is small enough, the
presence of just one polariton can block the resonant injection of
an additional polariton, because the polariton-polariton
interaction can shift the resonance frequency by an amount of the
order of the linewidth or even larger. The emitted light is
therefore strongly anti-bunched. If a pulsed pump is used, this
may result in a single-photon light source. The quantum polariton
blockade effect here considered is reminiscent of the one proposed
for atomic matter waves~\cite{Iacopo_atom} and, more closely, for
photons in cavities with a nonlinear atomic
medium~\cite{Imamoglu,Kimble}. An important advantage of using
polaritons comes from the strong exciton-exciton interactions,
while the photonic component guarantees an efficient and fast
coupling to the radiative modes outside the cavity where the
emission takes place.

The quantum emission properties of the proposed system are
quantitatively studied by means of the master equation for the
coupled exciton and cavity photon fields including losses. In
particular, we have studied the behavior of the second-order
coherence function $g^{(2)}$ as a function of the relevant
physical parameters, and we have identified the regimes where the
antibunching is most effective. These results are then used to
characterize the emission in the presence of a pulsed source,
which is shown to provide a train of single-photon pulses.

We start our theoretical treatment by recalling the quantum
Hamiltonian model
\cite{Iac_Crist_1,review_Deveaud,review_Cristiano} describing a
quantum well exciton strongly coupled to a planar microcavity
photon mode, namely
\begin{eqnarray}
H &=& \int d\vec{x} \sum_{i,j \in \{X,C\} }\hat{\Psi}_i^\dag
(\vec{x}) h_{i,j}^0(-i\nabla) \hat{\Psi}_j (\vec{x}) \nonumber
\\&&
+ \frac{\hbar \kappa}{2} \int d\vec{x} \hat{\Psi}_X^\dag (\vec{x})\hat{\Psi}_X^\dag (\vec{x})\hat{\Psi}_X (\vec{x})\hat{\Psi}_X (\vec{x}) \nonumber
\\ &&
 - \frac{\hbar \Omega_R}{n_{sat}}\int d\vec{x}\hat{\Psi}_C^\dag (\vec{x})\hat{\Psi}_X^\dag (\vec{x})\hat{\Psi}_X (\vec{x})\hat{\Psi}_X (\vec{x})
 + h.c \nonumber
 \\&&
 + \int d\vec{x} \hbar F_p(\vec{x},t) e^{-i\omega_p t }\hat{\Psi}_C^\dag(\vec{x})
 + h.c ~,
\end{eqnarray}
where the field operators $\hat{\Psi}_{X,C}$ describe excitons
($X$) and cavity photons ($C$). These operators depend on the
in-plane position wavevector $\vec{x}$, which is perpendicular to
the growth direction $z$. They satisfy Bose commutation rules
$[\hat{\Psi}_i (\vec{x}) ,\hat{\Psi}^\dag_j (\vec{x'})] =
\delta^2( \vec{x}-\vec{x'}) \delta_{i,j}$. The linear term,
including the exciton and planar microcavity photon kinetic energy
(the motion along $z$ is quantized), reads
\begin{equation}
\label{linear}
h^0(-i\nabla) = \hbar \left( \begin{array}{cc}{} \omega_X(-i\nabla)  & \Omega_R \\
\Omega_R & \omega_C(-i\nabla) + V_C(\vec{x})  \end{array} \right)~,
\end{equation}
where the exciton-photon coupling, responsible for the appearance
of the polariton eigenmodes, is quantified by the vacuum Rabi
frequency $\Omega_R$. $V_C(\vec{x})$ describes the photonic dot
confining potential due to the lithographic patterning. Two
contributions are responsible for the polariton nonlinearities,
namely the exciton-exciton interaction (modeled through a
repulsive contact interaction potential with strength $\hbar
\kappa$) and the anharmonic exciton-photon coupling (depending on
the exciton oscillator strength saturation density $n_{sat}$).
Finally, $F_p(\vec{x},t)$ describes the applied pump field with
frequency $\omega_p$.
\begin{figure}[t!]
\begin{center}
\includegraphics[width=8cm]{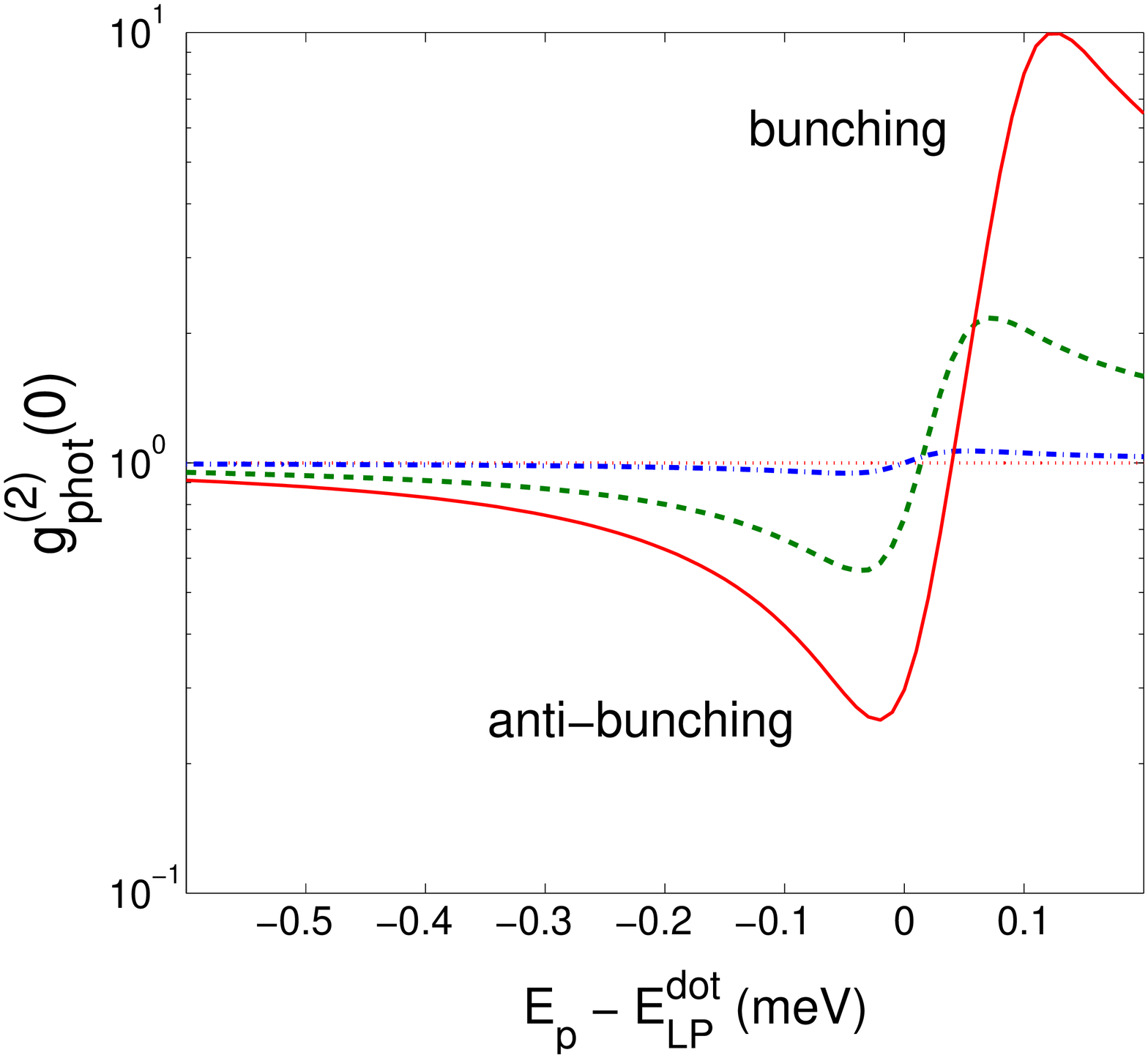}
\caption{Second order coherence function $g^{(2)}(0)$ versus pump
detuning $\hbar \omega_p - \hbar \omega_{LP}^{dot}$ (meV) for
three different cavity-exciton detunings. Solid, dashed,
dotted-dashed line: $\hbar(\omega_C^{dot} - \omega_X) =$ 5, 0, -5
meV. Parameters: $\hbar\omega_{nl} = $ 0.4 meV, $\hbar \gamma_X =
\hbar \gamma_C = $ 0.1 meV and continuous wave excitation field
$\hbar\mathcal{F}_0 = 10^{-2}$ meV. }\label{frequency_dependance}
\end{center}
\end{figure}

The photon field operator can be expanded in terms of the confined
modes in the photonic dot, namely $\hat{\Psi}_C (\vec{x}) = \sum_j
\phi_{C,j}(\vec{x}) ~\hat{a}_j$, where $\phi_{C,j}$ is the
normalized wavefunction of the $j$-th mode of energy $\hbar
\omega^{dot}_{C,j}$, while $\hat{a}_{j}$ is the corresponding
annihilation operator. Since the exciton kinetic energy is
negligible compared to the photonic one (i.e., the exciton mass
can be approximated as infinite), it is convenient to use  the
same basis to expand the exciton operator as$\hat{\Psi}_X(\vec{x})
= \sum_j \phi_{C,j}(\vec{x}) ~\hat{b}_j$, where $\hat{b}_j$ is the
corresponding exciton annihilation operator. In fact, it can be
easily seen from Eq. (\ref{linear}) that each photon mode is
coupled only to the exciton mode with the same spatial
wavefunction, implying that the polariton eigenmodes have the same
spatial wavefunction as the photonic dot modes. In the following,
we will be interested in studying the dynamics of the fundamental
photonic mode confined in the photonic dot, when this is close to
resonance with the exciton level. In the case of a strong photonic
confinement, the energy spacing between confined photon modes can
become much larger than the mode spectral linewidth and the energy
detuning between the quantum well exciton resonance and the
considered photon mode. In this limit and for quasi-resonant
excitation, we can safely simplify our quantum description by
retaining in the Hamiltonian only the fundamental photonic dot
mode of energy $\hbar \omega_{C}^{dot}$ and the exciton mode
having the same spatial wavefunction $\phi_{C}(\vec{x})$.
\begin{figure}[t!]
\begin{center}
\includegraphics[width=8cm]{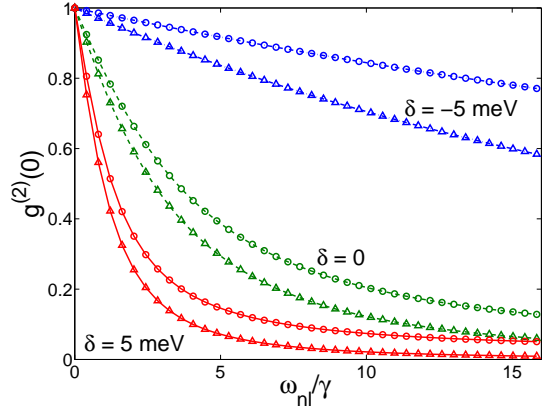}
\caption{Second order coherence function $g^{(2)}(0)$ for the
photon (circles) and exciton (triangles) as a function of the
normalized nonlinear coefficient $\omega_{nl}/\gamma$ for three
different values of $\delta = \hbar(\omega_C^{dot} - \omega_X)$. Parameters: $\hbar
\gamma_X = \hbar \gamma_C  $ = 0.1 meV and $\omega_{LP}^{dot} =
\omega_{p}$.} \label{g0_dependance}
\end{center}
\end{figure}
Thus, in the following, we will consider the following effective
Hamiltonian:
 \begin{eqnarray}\label{Ham_monomode}
H_{eff} &=& \hbar \omega_X b^\dag b +\hbar \omega_C^{dot} a^\dag a
+ \hbar \Omega_R b^\dag a + \hbar\Omega_R b a^\dag  \nonumber
\\&& + \frac{\hbar \omega_{nl}}{2} b^\dag b^\dag b
b-\alpha_{sat} \hbar \Omega_R b^{\dag}b^{\dag} a b -
\alpha_{sat} \hbar \Omega_R a^{\dag} b^{\dag} b b \nonumber
\\&&
+ \hbar \mathcal{F}_0(t) e^{-i\omega_p t }a^\dag + \hbar
\mathcal{F}_0^*(t) e^{i\omega_p t }a ~,
\end{eqnarray}
where $a$ and $b$ are the annihilation operators of the considered
photonic dot and exciton mode respectively. The parameters
involved in the effective Hamiltonian are the applied laser
amplitude $ \mathcal{F}_0(t) = \int d\vec{x} F_p(\vec{x},t)
{\phi_C}^*(\vec{x})$, while $\alpha_{sat} = \frac{1}{n_{sat}}\int
d\vec{x}  |{\phi_C}(\vec{x})|^4 $ and $\omega_{nl}= \kappa \int d
\vec{x}  |\phi_C(\vec{x})|^4$ are the effective nonlinear
coefficients. The saturation coefficient $\alpha_{sat}$ will be
neglected in the numerical solution because $\frac{\alpha_{sat}
\hbar \Omega_R}{\hbar \omega_{nl}/2} = \frac{2
\Omega_R}{n_{sat}\kappa} \ll  1$ for typical III-V microcavity
parameters\cite{review_Cristiano}.

To give the dependance of the nonlinear coefficient $\omega_{nl}$
on the photonic dot confinement, we have considered two simple
geometries with infinite confinement barriers. In the case of a
squared dot, the normalized wavefunction is $\phi(x,y) =
\frac{2}{L}\sin (\frac{\pi}{L} x) \sin (\frac{\pi}{L} y) $ where
$L$ is the lateral size. In the cylindrical case, $\phi(r)
=\frac{1.087}{R} J_0 (2.405 \, r/R)$ where $R$ is the radius of
the cylinder and $J_0$ the zero-th order Bessel function. The
values of the geometric coefficients are $\int_{square} d \vec{x}
|\phi_C(\vec{x})|^4 = 2.25 /L^2$, and $\int_{cylinder} d \vec{x}
|\phi_C(\vec{x})|^4 = 2.67 /(2R)^2$, showing the inverse
proportionality between $\omega_{nl}$ and the lateral area of the
photonic mode.
\begin{figure}[t!]
\begin{center}
\includegraphics[width=8.5cm]{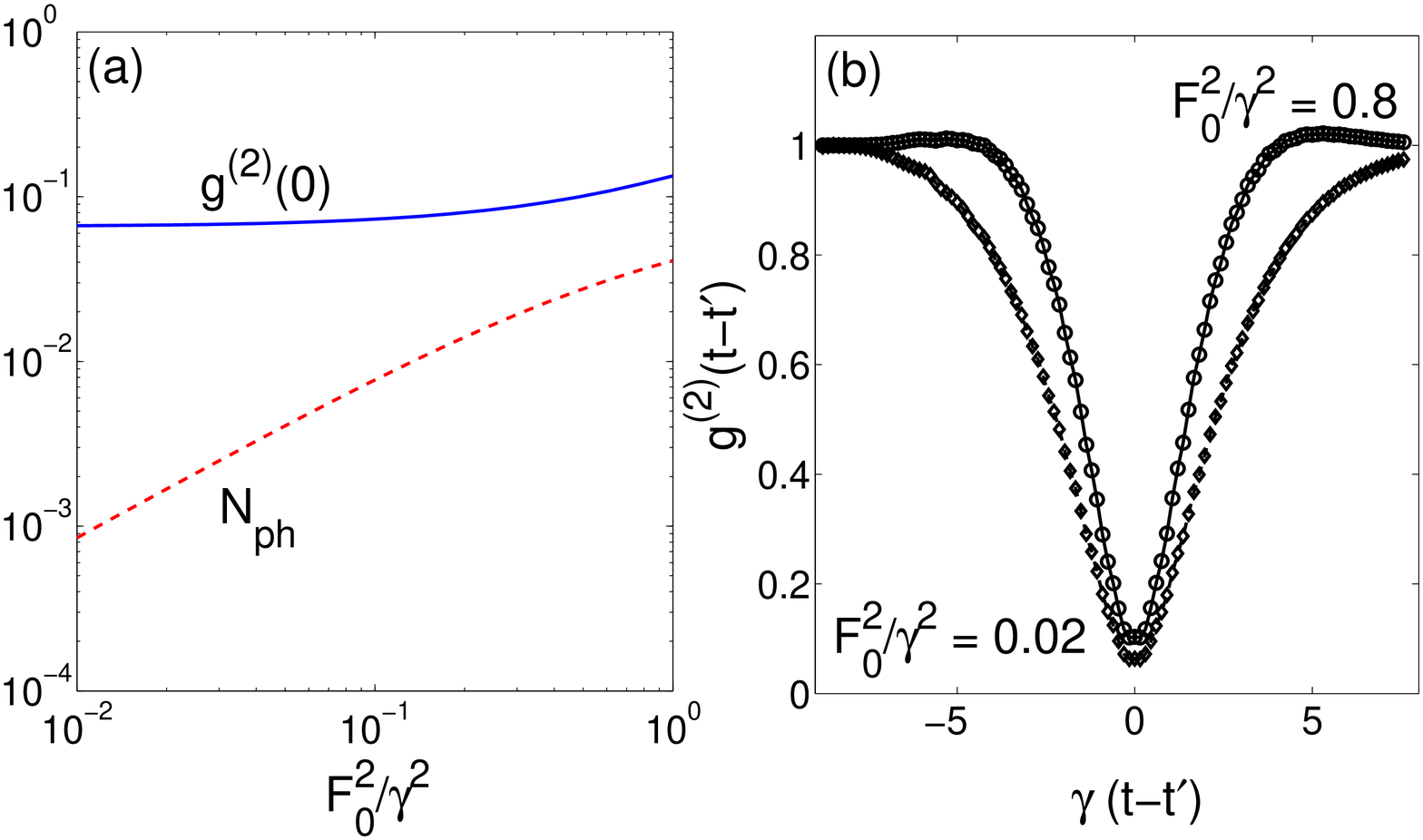}
\caption{(a) Second order coherence function $g^{(2)}_{phot}(0)$
(solid line) and intracavity photon population (dotted line)
inside the dot as a function of pump power in the continuous wave
regime. (b) Second order coherence function $g^{(2)}_{phot}(t,t')$
at a fixed $t'$ as a function of time for two different pump
powers. Parameters: $\hbar(\omega_C^{dot} -\omega_X)= $ 5 meV,
$\hbar\omega_{nl} = $
 1 meV, $\hbar \gamma_X = \hbar \gamma_C = $ 0.1 meV.}
\label{power_g2}
\end{center}
\end{figure}
In order to study the quantum dynamics, it is convenient
to work in the rotating frame described by the operator $R =
e^{i(\omega_p t (a^\dag a  + b^\dag b))}$. The rotating frame Hamiltonian is:
\begin{eqnarray}
\tilde{H}_{eff} &=& \hbar \Delta\omega_X b^\dag b +\hbar
\Delta\omega_C a^\dag a + \hbar \Omega_R b^\dag a + \hbar\Omega_R
b a^\dag \nonumber \\&& + \frac{\hbar \omega_{nl}}{2} b^\dag
b^\dag b b + \hbar \mathcal{F}_0(t) a^\dag + \hbar
\mathcal{F}_0^*(t) a ~,
\end{eqnarray}
where $\Delta\omega_C = \omega_C^{dot} - \omega_p$ and
$\Delta\omega_X = \omega_X - \omega_p$ respectively. To describe
the quantum dynamics in presence of damping, we have considered
the master equation for the density matrix $\rho(t)$:
\begin{eqnarray}\label{rho_evolution_equation}
\frac{\partial\tilde{\rho}}{\partial t} &=
&\frac{i}{\hbar}[\tilde{\rho},\tilde{H}_{eff}] + \gamma_C
(a\tilde{\rho} a^\dag - 1/2(a^\dag a \tilde{\rho} + \tilde{\rho}
a^\dag a)) \nonumber\\&& + \gamma_X  (b\tilde{\rho} b^\dag -
1/2(b^\dag b \tilde{\rho} + \tilde{\rho} b^\dag b)) ~,
\end{eqnarray}
where $\tilde{\rho} = R\rho R^\dag$, while $\gamma_X$ and
$\gamma_C$ are the homogeneous broadening of the exciton and
photon modes. The master equation can be solved by expanding the
density matrix over a Fock basis, namely
\begin{equation}
\tilde{\rho}(t)=\sum_{n_X',n_C',n_X,n_C}
\tilde{\rho}_{n_X',n_C',n_X,n_C}(t) |n_X',n_C'\rangle \langle
n_X,n_C| ~,\end{equation} where $n_X$ and $n_C$ are the number of
excitons and photons respectively. In the following, we will be
interested in the two-time second-order coherence function
\cite{Walls}, defined as: \onecolumngrid
\begin{eqnarray}\label{g_2(0)}
g^{(2)}_{phot}(t,t') &  = & \frac{G^{(2)}_{phot}(t,t')}{N_{ph}(t)
N_{ph}(t') } =  \frac{ Tr\left({a} \ \mathcal{U}_{t,t'} \left[{a}
\tilde{\rho}(t') {a}^\dag\right] {a}^\dag \right)}{Tr( {a}
\tilde{\rho}(t) {a}^\dag) Tr({a} \tilde{\rho}(t') {a}^\dag )}
 = \frac{\sum_{m,n} m \theta_{n,m,n,m}(t,t')}
{\sum_{m,n} m \tilde{\rho}_{n,m,n,m}(t) \sum_{m,n} m \tilde{\rho}_{n,m,n,m}(t')}~,\ \forall t>t' \nonumber\\
\theta(t,t') &=& \mathcal{U}_{t,t'}\left[\sum_{n,m,n',m'}\tilde{\rho}_{n',m',n,m}(t') \sqrt{m m'} |n',m'-1\rangle \langle n,m-1|\right]~,
\end{eqnarray}
\twocolumngrid where $\mathcal{U}_{t,t'}$ is the evolution
superoperator associated to the master equation
(\ref{rho_evolution_equation}).

As we have already discussed, we will consider the case of an
applied optical field with frequency $\omega_p$ close to the
frequency $\omega_{LP}^{dot} = \frac{\omega_C^{dot} + \omega_X}{2}
 - \sqrt{\Omega_R^2 + \frac{(\omega_C^{dot} - \omega_X)^2}{4}}$ of the fundamental confined
polariton mode in the dot. In the case of a continuous wave
excitation, we give in Fig. \ref{frequency_dependance} an example
of the dependence of the equal-time second-order coherence
$g^{(2)}_{phot}(0) \equiv g^{(2)}_{phot}(t,t)$ on the pump
detuning $\omega_p - \omega_{LP}^{dot}$ for a given set of
parameters. For a pump laser frequency red-detuned or close to
resonance with the fundamental polariton resonance,
$g^{(2)}_{phot}(0) < 1$, implying sub-poissonian statistics and
antibunching. This is the regime where the polariton quantum
blockade is working. Indeed, the photon injection is inhibited
when only a small number of polaritons are already inside the dot
due to the interaction-induced blueshift of the polariton
resonance. For large values of $\omega_{nl}/\gamma$, only one
polariton can be present in the dot with a vanishing probability
of having two at the same time, implying $g^{(2)}_{phot}(0)
\approx 0$. On the other-hand, for a blue-detuned laser,
$g^{(2)}_{phot}(0)
> 1$, implying bunching.
In Fig. \ref{frequency_dependance}, there are three curves
corresponding to different detunings $\delta = \hbar(
\omega^{dot}_C - \omega_{X})$. Since the nonlinearity is due to
the excitonic fraction of the polaritonic mode, the photonic
antibunching is more pronounced for $\delta > 0$ . As shown in
Fig. \ref{g0_dependance}, the minimum value of $g^{(2)}_{phot}(0)$
depends critically on the ratio $\omega_{nl}/\gamma$, where the
polariton mode linewidth $\gamma = |X_{LP}|^2 \gamma_X +
|C_{LP}|^2 \gamma_C$, being $|X_{LP}|^2$ and $|C_{LP}|^2$ the
excitonic and photonic fractions of the lower polariton mode
respectively. Here, for the sake of clarity, we have performed
calculations taking $\gamma_X = \gamma_C = \gamma$. The
antibunching behavior ($g^{(2)}_{phot}(0) < 1)$ starts to be
significant when $\omega_{nl}/\gamma \sim 1$. In order to have
$\omega_{nl}/\gamma =1$ , with a polariton linewidth $\hbar \gamma
= 0.1$ meV and with a realistic nonlinear
coefficient\cite{review_Cristiano,review_Deveaud} $\hbar\kappa =
1.5 \times 10^{-2} (\mu {\rm m})^2$ meV (corresponding to an
exciton blueshift of 0.15 meV in presence of $10^9$ cm$^{-2}$
excitons), a cylindrical dot with diameter $2R = 0.67$ $\mu m$
would be required. Reducing further the size allows one to enter
the strong quantum blockade regime $\omega_{nl}/\gamma \gg 1$. For
example, using the same nonlinear coefficient, a square dot with
lateral size $L = 0.2$ $\mu m$ gives $\omega_{nl}/\gamma = 8.4$.
In general,
 there is slight asymmetry between the photonic and excitonic antibunching
($g_{phot}^{(2)}(0) > g_{exc}^{(2)}(0)$) even at zero detuning.
This asymmetry occurs because the nonlinearity is due to the
exciton.
\begin{figure}[t!]
\begin{center}
\includegraphics[width=8cm]{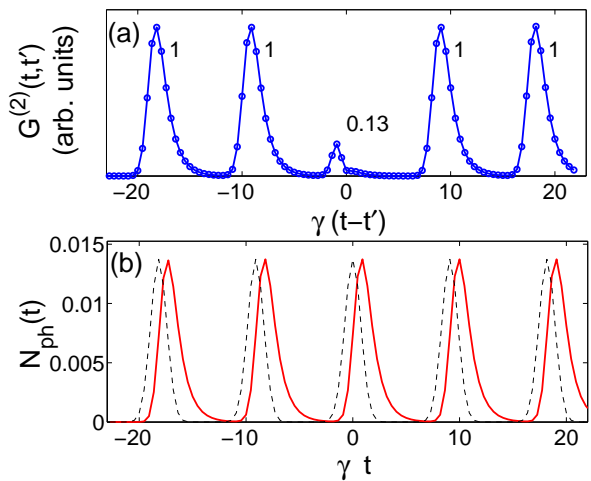}
\caption{(a) Second order correlation function
$G^{(2)}_{phot}(t,t')$ for $t' = 6$ ps under the excitation of a
train of gaussian pulses (pulse duration of 5 ps) separated from
each other by 60 ps. The normalized area is indicated for each
peak. (b) Corresponding intra-cavity photon population
$N_{ph}(t)$.
 Dotted line: shape of the pump amplitude $\mathcal{F}_0(t)$.
 Parameters: $\hbar(\omega_C^{dot}-\omega_X)= $ 5 meV,
 $\hbar(\omega_p-\omega_{LP}^{dot})= $ 0.05 meV,
 $\hbar\omega_{nl} = $ 1.1 meV, $|{\mathcal F_0}|^2/\gamma^2 = 0.09$,
 $\hbar\gamma_X = \hbar\gamma_C = $ 0.1 meV,
 so $1/\gamma =$ 6.6 ps.}\label{pulse_g2}
\end{center}
\end{figure}

Given the current interest for quantum devices, it is useful to
characterize the peculiar figures of merit of the single-photon
source under consideration. To have an efficient single-photon
source, we need to maximize the photon population $N_{ph}$,
keeping the sub-poissonian character strong. To address this
issue, in Fig. \ref{power_g2}(a) we have plotted
$g^{(2)}_{phot}(0)$ and the intra-cavity photon population
$N_{ph}$ as a function of the normalized incident intensity
$|{\mathcal F_0}|^2/\gamma^2$ of the cw laser. For $|{\mathcal
F_0}|^2/\gamma^2 \to 0$, $g^{(2)}_{phot}(0)$ asymptotically
converges to a minimum value, but the population $N_{ph}$ goes to
0. For increasing $|{\mathcal F_0}|^2/\gamma^2$, $N_{ph}$
increases, but $g^{(2)}_{phot}(0)$ eventually grows up. For the
parameters here used, the crossover occurs for $N_{ph} \approx
0.01$. In Fig. \ref{power_g2}(b), the dependence of
$g^{(2)}_{phot}(t,t')$ on the relative time $t-t'$ is shown for
two excitation intensities. It is apparent that the temporal width
of the antibunching dip is directly related to the inverse
polariton linewidth $1/\gamma$, at least in the limit $|{\mathcal
F}_0|^2/\gamma^2 \to 0$.

Since the polariton quantum blockade effect relies on the resonant
character of the excitation, one can wonder whether the effect is
robust even in the pulsed excitation regime. In addition to the
strong sub-poissonian photon statistics, the efficiency and the
repetition rate  are the relevant quantities in the pulsed
excitation case. We have solved the dynamics using a train of
excitation pulses and we have found that by using Fourier-limited
pulses with spectral linewidth comparable to the polariton one and
a repetition rate $\Gamma \ll \gamma$ , the suppression of the
two-photon probability approaches the cw case. As an illustrative
example, in Fig. \ref{pulse_g2}(a) we show the time-dependent
second-order correlation function $G^{(2)}_{phot}(t,t')$. The
depletion of the central peak (which would not occur for a source
with poissonian statistics, such as an attenuated laser beam)
demonstrates the strong single-photon character of the present
source even in the pulsed regime. The quantity $\eta = \gamma_C
\int_{\Delta T} N_{ph}(t) dt$ (where $\Delta T$ is the time
interval between two consecutive pulses) represents the averaged
number of photons emitted per pulse. As shown in Fig.
\ref{pulse_g2}(b), a repetition rate $\Gamma = \gamma/10$ is
enough to avoid pulse overlap. The effective quantum bit exchange
rate of the present quantum source would be $r = \eta \Gamma$.
With respect to the example in Fig. \ref{pulse_g2}(b), we have
$\eta \simeq 0.01$, implying $r \simeq 0.1$ GHz.

In conclusion, the present work has predicted the rich quantum
nonlinear dynamics of a quantum well exciton transition strongly
coupled to a photonic quantum dot mode, showing the potential for
the realization of a single-photon source with controllable
properties based on the polariton quantum blockade effect.

We thank G. Bastard, B. Deveaud, C. Diederichs, G. Dasbach, O. El
Da\"{\i}f, I. Favero, J.M. G\'{e}rard, F. Morier-Genoud, A.
Imamo\u{g}lu, N. Regnault, J. Tignon for discussions. LPA-ENS is a
"Unit\'{e} Mixte de Recherche Associ\'{e} au CNRS (UMR 8551) et
aux Universit\'{e}s Paris 6 et 7".

\end{document}